# Strain-Stabilized Interfacial Polarization Tunes Work Function Over 1 eV in RuO$_2$/TiO$_2$ Heterostructures

Seung Gyo Jeong[1,*], Bonnie Y.X. Lin[2], Mengru Jin[3], In Hyeok Choi[4], Seungjun Lee[5], Zhifei Yang[1, 6], Sreejith Nair[1], Rashmi Choudhary[1], Juhi Parikh[1], Anand Santhosh[1], Matthew Neurock[1], Kelsey A. Stoerzinger[1], Jong Seok Lee[4], Tony Low[5], Qing Tu[3], James M. LeBeau[2], and Bharat Jalan[1,*]

[1]Department of Chemical Engineering and Materials Science, University of Minnesota−Twin Cities, Minneapolis, Minnesota 55455, USA

[2]Department of Materials Science and Engineering, Massachusetts Institute of Technology, Cambridge, Massachusetts 02139, USA

[3]Department of Materials Science and Engineering, Texas A&M University, College Station, Texas 77843, USA

[4]Department of Physics and Photon Science, Gwangju Institute of Science and Technology (GIST), Gwangju 61005, Republic of Korea

[5]Department of Electrical and Computer Engineering, University of Minnesota, Minneapolis, Minnesota 55455, USA

[6]School of Physics and Astronomy, University of Minnesota, Minneapolis, Minnesota 55455, USA

[*]Corresponding authors: jeong397@umn.edu; bjalan@umn.edu

**Abstract**

Interfacial polarization—charge accumulation at the heterointerface—is a well-established tool in semiconductors, but its influence in metals remains unexplored. Here, we demonstrate that interfacial polarization can robustly modulate surface work function in metallic rutile $RuO_2$ layers in epitaxial $RuO_2/TiO_2$ heterostructures grown by hybrid molecular beam epitaxy. Using multislice electron ptychography, we directly visualize polar displacements of transition metal ions relative to oxygen octahedra near the interface, despite the conductive nature of $RuO_2$. This interfacial polarization enables over 1 eV modulation of the $RuO_2$ work function, controlled by small thickness variations (2–3 nm), as measured by Kelvin probe force microscopy, with a critical thickness of ~4 nm—corresponding to the transition from fully-strained to relaxed film. These results establish interfacial polarization as a powerful route to control electronic properties in metals and have implications for designing tunable electronic, catalytic, and quantum devices through interfacial control in polar metallic systems.

**Main**

Interfacial polarization is a cornerstone of modern oxide electronics, playing a pivotal role in phenomena ranging from two-dimensional electron gases at perovskite interfaces[1] to emergent ferroelectricity in confined geometries[2]. In insulating or semiconducting oxides, such polarization can be engineered via epitaxial strain, chemical modulation, or structural asymmetry, leading to functionalities such as switchable dipoles, band bending, and interface conductivity[3, 4, 5]. In contrast, the extension of these design principles to metallic systems has long been thought untenable due to the strong screening of long-range Coulomb interactions by free carriers[6]. This view has been upended by the emerging class of polar metals[7, 8], in which metallic conductivity and non-centrosymmetric, polar structures coexist—a concept once considered incompatible. Experimental and theoretical studies have shown that polar distortions can persist in metals under specific symmetry, bonding, and lattice-dynamical conditions[9, 10, 11, 12]. Notably, chemically doped ferroelectrics[4, 13], strain-tuned perovskites[14], and symmetry-driven design strategies[15] have yielded polar metals exhibiting phenomena such as enhanced thermoelectricity[10, 16, 17], nontrivial band topology[18, 19], spin-polarized transport[20, 21], and ferroelectric-like superconductivity[22, 23]. These discoveries challenge long-standing assumptions and suggest a broader design space for materials that harness both conductivity and polarity.

Several 3$d$ perovskite-based systems have demonstrated coexisting polar displacements, metallicity, and even magnetism[24, 25, 26]. However, exploration beyond perovskites remains limited, and interfacial polarization in metallic rutile oxides—a structurally distinct family—has not been previously demonstrated. Furthermore, direct, spatially resolved identification of such polarization in metallic systems has not been possible thus far, hindering our understanding of its microscopic origin and macroscopic impact.

In this Article, we investigate epitaxial heterostructures of $RuO_2$ on $TiO_2$ (110) to reveal the existence and consequences of interfacial polarization in a rutile-based metallic system. $RuO_2$ is a chemically stable, highly conductive oxide with a tetragonal rutile structure, widely studied for its catalytic properties but not known to exhibit polar distortions. Using hybrid molecular beam epitaxy (MBE), we synthesize atomically smooth, single-crystalline $RuO_2$ films and examine their interfacial structure and electronic properties through multislice electron ptychography, Kelvin probe force microscopy (KPFM), and electrical transport. We uncover a robust, strain-induced interfacial polarization at the buried interface, which induces an electrical dead layer and enables modulation of the surface work function by over 1 eV—an extraordinary large effect in a metallic system. Our findings establish a new paradigm that goes beyond the mere observation of interface dipoles in metals: (1) interfacial polarization in metallic layers can be stabilized and controlled via epitaxial strain and heterostructure, and (2) this polarization can significantly modulate the work function and electronic transport properties of the metallic layers. Together, these insights provide a powerful platform for designing functional materials in which electric fields, mechanical strain, or thickness serve as tunable parameters to control both structural and electronic responses. This work opens new frontiers in polar metals and interface-driven functionalities, with implications for oxide electronics, spintronics, and energy applications.

**Epitaxial, atomically precise $RuO_2$ heterostructures via hybrid molecular beam epitaxy**

To probe the presence of interfacial polarization in metallic $RuO_2$, we synthesized high-quality $TiO_2/RuO_2/TiO_2$ (110) heterostructures using hybrid MBE. For $RuO_2/TiO_2$ (110), the lattice mismatches between $RuO_2$ and $TiO_2$ are −4.7% along the [001] and +2.3% along the [1$\bar{1}$0]. Specifically, a 2.5 nm $TiO_2$ buffer layer, a 3.8 nm thick $RuO_2$ film, and a 2.5 nm $TiO_2$ capping layer were sequentially grown on a $TiO_2$ (110) single-crystalline substrate. The symmetric architecture—with identical buffer and cap layers—was designed to eliminate surface- and substrate-induced variations, thereby isolating interfacial effects intrinsic to the $RuO_2$/$TiO_2$ interfaces. X-ray reflectivity (XRR) measurements (Fig. 1a, scattered symbols) show pronounced Kiessig fringes, indicating well-defined layer thicknesses and interfaces. The excellent agreement between the experimental data and the fitting model (solid line in Fig. 1a) confirms the precise control of the $RuO_2$ thickness, $t_{RuO_2}$ = 3.8 nm. Complementary structural characterization using X-ray diffraction (XRD) $\theta$-$2\theta$ scans (Fig. 1b) shows strong (110) Bragg reflections consistent with the rutile phase, accompanied by Laue oscillations near the $TiO_2$ (110) peak. These features signify a high-quality heteroepitaxy. An atomic force microscopy (AFM) image (inset, Fig. 1b) further verifies an atomically flat surface morphology, with well-defined step-terrace structures and a root mean square roughness ($S_q$) of 155 pm.

**Atomic-scale visualization of polar displacements via electron ptychography**

The presence of polar displacements at the $RuO_2$/$TiO_2$ interfaces was directly visualized using cross-sectional multislice electron ptychography (Figs. 1c–1e and Extended Data Figs. 1–3), which revealed the emergence of out-of-plane electrical dipoles. Multislice electron ptychography[27, 28] is a computational phase retrieval technique that uses coherent interference in four-dimensional scanning transmission electron microscopy (4D-STEM) to iteratively reconstruct the incident electron probe and the phase shifts introduced by electrons scattering within the object, which scales with the projected potential. The result is to largely remove the effects of dynamical scattering and the deconvolved object with 3D information[28]. The greatly improved resolution and linear atomic number dependence allow for precise measurement of both oxygen and cation atom column positions, and thus polar displacements, *within* the electron microscopy sample.

To ensure accurate measurement of interfacial polar distortions, we focused on the $TiO_2$/$RuO_2$/$TiO_2$ heterostructure, which symmetrically preserves the ultrathin $RuO_2$ layer. In contrast, samples $RuO_2$ as the topmost layer—such as those in $RuO_2$/$TiO_2$ stacks without a TiO$_2$ capping layer—are susceptible to surface damage during sample preparation for electron microscopy.

The left panel of Fig. 1c displays a high-resolution ptychographic reconstruction of the TiO$_2$/RuO$_2$/TiO$_2$ (110) heterostructure. The reconstructed object consisted of 20 slices, each 1 nm thick. The average of only the middle three slices (the 9th-11th slice) is displayed and analyzed. This removes the electron beam-induced damage and amorphous sample contamination on the surface, and also reduces the effect of tilt and surface relaxation of the thin sample. The Ru, Ti, and O atom columns in the rutile structure are clearly resolved in the averaged reconstructed phase. Since the reconstructed phase scales with atomic number ($Z^{0.67}$) [28], RuO$_2$/TiO$_2$ interfaces were identified by analyzing the relative cation peak phase (Extended Data Fig. 2), marked by horizontal lines. These interfaces are atomically abrupt, consistent with XRR results. Based on the boundaries obtained from the ptychographic reconstruction, $t_{RuO_2}$ is determined to be 3.9 nm, in excellent agreement with the XRR analysis (3.8 nm). With precise measurement of the Ru, Ti, and O atom column positions, we extracted displacement vectors of Ru and Ti cations relative to their surrounding anion framework of four oxygens, shown as yellow arrows in the right panel of Fig. 1c. These displacements indicate the presence of polar distortions across both interfaces. Notably, the polarization vectors at the upper and lower interfaces point in opposite directions, each oriented from the TiO$_2$ into the RuO$_2$, consistent with mirror-symmetric interfacial polar fields.

To quantify this effect, Fig. 1d plots the average out-of-plane polar displacement component along a selected atom plane (marked in Fig. 1c). A significant out-of-plane polar displacement—up to ~9 pm (17.4 pm locally)—is observed in the TiO$_2$ layers and propagates into the adjacent RuO$_2$ regions, indicating a breakdown of inversion symmetry within the metallic RuO$_2$ at the interfaces. In contrast, in-plane displacements (Extended Data Fig. 3) are relatively small, emphasizing the anisotropic nature of the polar distortion. This observation aligns with first-principles predictions: although strained $RuO_2$ shows lattice instabilities at the Brillouin zone boundary[29], it lacks zone-center soft modes, suggesting strain alone cannot stabilize a polar ground state. We hypothesize that the spontaneous polarization arises from the coherently strained $TiO_2$ layer at the $RuO_2/TiO_2$ interface and extends into the $RuO_2$ layer, causing local symmetry breaking along the out-of-plane [110] direction. Although bulk $TiO_2$ is nominally non-polar, interfacial effects in the fully strained $RuO_2/TiO_2$ system—such as inversion symmetry breaking and the continuity of octahedral networks across the interface—can induce polar behavior. This coupling between polar $TiO_2$ and metallic $RuO_2$ layers underpins the emergence of interfacial polarization in this system.

**Thickness- and Strain-modulation of work-function on metal surface**

It is conceivable that the observed interfacial polarization, oriented from the $TiO_2$ substrate toward the $RuO_2$ film as revealed by electron ptychography (Fig. 1), should enhance the surface work function ($\Phi$) of $RuO_2$ by contributing an additional internal electric field at the interface. To test this hypothesis, we performed KPFM measurements on uncapped $RuO_2/TiO_2$ (110) thin films with varying $t_{RuO_2}$ (see Extended Data Fig. 4 for sample details). The KPFM setup, illustrated in Fig. 2a, measures the local contact potential difference ($V_{CPD}$) between the conductive tip and the sample surface, from which $\Phi$ is determined using $\Phi = \Phi_{tip} - qV_{CPD}$, where $\Phi_{tip}$ is the work function of the tip (calibrated by Ultraviolet Photoelectron Spectroscopy (UPS), see Methods for details) and $q$ is the elementary charge. Surface potential mapping for samples with $t_{RuO_2}$ = 3.1, 3.9, and 5.2 nm (Extended Data Figs. 5 and 6) revealed consistent topography with $S_q$ values around 200 pm and uniform $V_{CPD}$ across the scanned area. The small standard deviation in $V_{CPD}$ (~1.6 mV) ensured high precision, while the mean $V_{CPD}$ values showed significant dependence on $t_{RuO_2}$.

Notably, $\Phi$ exhibited a non-monotonic trend with a peak at a critical thickness ($t_c$) of ~ 4 nm, implying additional polarization-related effects. As shown in Fig. 2b, $\Phi$ increases from 4.246 eV at $t_{RuO_2}$ = 1.5 nm—close to the bulk $TiO_2$ value of 4.1 eV (Extended Data Fig. 7)[30]—to a maximum of 5.401 eV at $t_{RuO_2}$ = 3.5 nm. It then decreases up to $t_{RuO_2} \approx$ 7 nm, followed by a gradual linear increase, reaching 5.054 eV at 17.2 nm—consistent with previously reported values for (110)-oriented $RuO_2$ surfaces[31, 32]. This non-monotonic behavior deviates from the monotonic $\Phi$ vs. thickness trends observed in conventional heterostructures[33, 34, 35], and suggests the presence of an additional internal electric field, directed from $RuO_2$ toward the $TiO_2$ substrate. This electric field, originating from the polarization within $TiO_2$ and oriented toward $RuO_2$, effectively increases $\Phi$, consistent with the direction of polar displacements observed at the $RuO_2/TiO_2$ interface via electron ptychography (Fig. 1).

To understand the origin of the peak in $\Phi$ at a critical thickness of ~ 4 nm, we performed X-ray diffraction reciprocal space maps (RSMs) of $RuO_2/TiO_2$ (110) films, indicating that strain relaxation begins between 4 and 6.5 nm, aligning with the critical thickness observed in KPFM. RSMs around the $TiO_2$ (332) and (301) Bragg peaks (Figs. 2c, 2d, and Extended Data Fig. 4d) show that at $t_{RuO_2}$ = 4 nm, $RuO_2$ remains fully strained along [001], while partial relaxation appears at 6.5 nm as $Q_{001}$ shifts toward bulk $RuO_2$. The tensile strain along [1$\bar{1}$0] persists up to $t_{RuO_2}$ = 17 nm (Extended Data Fig. 4d), due to the smaller mismatch in that direction. This critical thickness also coincides with abrupt changes in optical second harmonic generation (SHG) symmetry with increasing film thickness—from polar *mm*2 to nonpolar 4/*mmm*—indicating a transition in the crystal symmetry of the $RuO_2$ layer[36]. We attribute the non-monotonic behavior of $\Phi$ to a strain-induced structural transition from a polar to a nonpolar phase in $RuO_2$, schematically depicted in Fig. 2e. In ultrathin films ($t_{RuO_2} < t_c$), the enhancement of $\Phi$ is consistent with increasing polar regions and a stronger surface electric field in the polar metallic RuO$_2$ phase. Above $t_c$, the formation of relaxed, nonpolar RuO$_2$ reduces the surface electric field, lowering Φ. These findings suggest that the electrical polarization, evolving with $t_{RuO_2}$, plays a central role in modulating the surface work function and that the origin of interfacial polarization is likely a result of the significant epitaxial strain in the strained RuO$_2$/TiO$_2$ (110) film.

Fig. 3 benchmarks our approach by comparing the modulated work functions of our RuO$_2$ films with those reported for a variety of material systems—including metal oxides[37, 38], elemental metals[38, 39, 40, 41], and two-dimensional materials[38, 42]—obtained via diverse methodologies in previous studies. Our films exhibit an exceptionally large work function modulation of ~1.15 eV (red line), far exceeding the typical variation (< 0.5 eV) achieved via thickness control of the metallic thin films (Extended Data Fig. 8) or external strain[43]. This variation is comparable to leading examples—such as the ~1.2 eV shifts observed in molecule-adsorption studies on metal surfaces[38] —yet is achieved here without any surface modification.

**Interfacial Polarization Effects on Electrical Transport**

To investigate the relationship between interfacial polarization and charge transport, we performed electrical transport measurements on RuO$_2$/TiO$_2$ heterostructures, as shown in Fig. 4, Extended Data Figs. 9 and 10. Figs. 4a and 4b display the room-temperature 2D conductivity ($\sigma_{2D}$) for RuO$_2$/TiO$_2$ and TiO$_2$/RuO$_2$/TiO$_2$ heterostructures as a function of $t_{RuO_2}$. Given the presence of interfacial polar displacements at the RuO$_2$/TiO$_2$ interface (Fig. 1) and the non-monotonic dependence of the work function on $t_{RuO_2}$ (Fig. 2b), it is plausible that the interfacial regions exhibit distinct transport behavior compared to the intrinsic bulk of RuO$_2$. Therefore, the measured $\sigma_{2D}$ arises from parallel conduction channels, as illustrated schematically in the insets of Figs. 4a and 4b. For the RuO$_2$/TiO$_2$ structure, $\sigma_{2D}$ can be modeled as a parallel sum of bulk conductivity ($\sigma_b$) and interfacial conductivity ($\sigma_{i1}$), expressed as: $\sigma_{2D} = (t_{RuO_2} - t_{i1}) \cdot \sigma_b +$ $t_{i1} \cdot \sigma_{i1} = t_{RuO_2} \cdot \sigma_b - t_{i1} \cdot (\sigma_b - \sigma_{i1})$, where $t_{i1}$ is the effective thickness of the interface. Assuming $\sigma_{i1} \ll \sigma_b$—consistent with prior reports of reduced conductivity in polar metals[8], this simplifies to: $\sigma_{2D} = t_{RuO_2} \cdot \sigma_b - t_{i1} \cdot \sigma_b$. In this linear form, the slope corresponds to $\sigma_b$, the x-intercept yields the effective interfacial thickness $t_{i1}$. Fitting the experimental data in Fig. 4a yields $\sigma_b = 15,875$ S/cm—comparable to the bulk RuO$_2$ value of 28,400 S/cm[44]—and $t_{i1} = 1.59$ nm. Using these results and a representative data point at $t_{RuO_2} = 2.1$ nm, we estimate $\sigma_{i1} = 4,117$ S/cm, supporting the self-consistency of the model and validating the assumption that $\sigma_{i1} \ll \sigma_b$. For TiO$_2$/RuO$_2$/TiO$_2$ heterostructures, the total $\sigma_{2D}$ includes contributions from both the lower and upper interfaces, expressed as: $\sigma_{2D} = t_{RuO_2} \cdot \sigma_b - (t_{i1} + t_{i2}) \cdot \{\sigma_b - (\sigma_{i1} + \sigma_{i2})\}$, where $t_{i2}$ and $\sigma_{i2}$ are effective thickness and conductivity at upper RuO$_2$/TiO$_2$ interface, respectively. Assuming $\sigma_{i1} + \sigma_{i2} \ll \sigma_b$, the simplified linear fit (Fig. 4b) gives $\sigma_b = 11,700$ S/cm and a combined interfacial thickness $t_{i1} + t_{i2} = 0.71$ nm. These linear trends persist at lower temperatures (1.8 K and 100 K), as confirmed in Extended Data Fig. 10, with fitting results summarized therein. Interestingly, both $\sigma_b$ and the effective interfacial thicknesses ($t_{i1}$ or $t_{i1} + t_{i2}$) increase as temperature decreases. The rise in $\sigma_b$ is indicative of metallic behavior, while the expansion of the interfacial thickness may reflect a broader interfacial region with suppressed conductivity. This temperature dependence supports the coexistence of polar and metallic phases within the RuO$_2$ layer.

Moreover, across all temperatures, $\sigma_b$ in TiO$_2$/RuO$_2$/TiO$_2$ heterostructures is consistently lower than in RuO$_2$/TiO$_2$ films, and $t_{i1} + t_{i2}$ is smaller than $t_{i1}$ alone. The extracted value of $t_{i1} + t_{i2} = 0.71$ nm at 300 K, or roughly two-unit cells of RuO$_2$, agrees well with the polar interfacial region observed via ptychography (Fig. 1c). This reduction in effective interfacial region suggests that the opposing interfacial polarizations at the top and bottom interfaces of the TiO$_2$/RuO$_2$/TiO$_2$ stack—confirmed by Fig. 1c—partially cancel each other, reducing the net interfacial region. Interfacial polarization effects are further reflected in the 2D carrier density ($n_{2D}$) and mobility ($\mu$) as functions of $t_{RuO_2}$, shown in Figs. 4c–f. These quantities were extracted from magnetic ($H$)-field-dependent Hall measurements at room temperature, using linear fits between ±9 T (Fig. 3d inset and Extended Data Fig. 9). For RuO$_2$/TiO$_2$ films, $n_{2D}$ shows a dip, while $\mu$ exhibits a peak near $t_{RuO_2} \approx 3.5$ nm—closely matching the $t_c$ observed in KPFM measurements. In TiO$_2$/RuO$_2$/TiO$_2$ films, although $n_{2D}$ remains relatively featureless (Fig. 4d), $\mu$ shows a peak at $t_{RuO_2} \approx 1.9$ nm, aligning with the smaller interfacial thickness ($t_{i1} + t_{i2}$) compared to RuO$_2$/TiO$_2$ films. While these measurements reveal a clear "anomaly" in both $n_{2D}$ and $\mu$ as a function of $t_{RuO_2}$, they are based on a simplified model that assumes single-channel conduction and neutral interfaces. This assumption overlooks the presence of an electric field in the interfacial region, which is expected to influence the charge conduction profile across the film thickness. A detailed treatment of these interfacial effects should be the focus of future investigations.

**Conclusions**

In summary, our study demonstrates the coexistence of interfacial polarization and metallicity in epitaxial RuO$_2$/TiO$_2$ heterostructures and their controllability via epitaxial design. Cross-sectional multislice electron ptychography reveals atomic-scale polar displacements directed from TiO$_2$ to RuO$_2$ at interfaces. Complementary KPFM measurements show a non-monotonic thickness dependence of the surface work function, which increases by more than 1 eV for RuO$_2$ thicknesses below ~4 nm—coinciding with the regime where epitaxial strain is preserved. Notably, the emergence of electrical polarization is intimately linked to the transport properties of ultrathin, metallic RuO$_2$. By accounting for interfacial contributions, we extract the intrinsic conductivity of RuO$_2$ layers in the heterostructure, finding values comparable to bulk RuO$_2$. This analysis further reveals that the thickness of an electrical dead layer is ~1.6 nm in RuO$_2$/TiO$_2$ structures, while it is reduced to ~0.7 nm in TiO$_2$/RuO$_2$/TiO$_2$ heterostructures, consistent with the opposing interfacial polarization observed via electron ptychography. These findings establish a pathway for realizing polar metallic states via interfacial design, offering new opportunities to simultaneously engineer polarization and conductivity in rutile oxide systems.

**Figures**

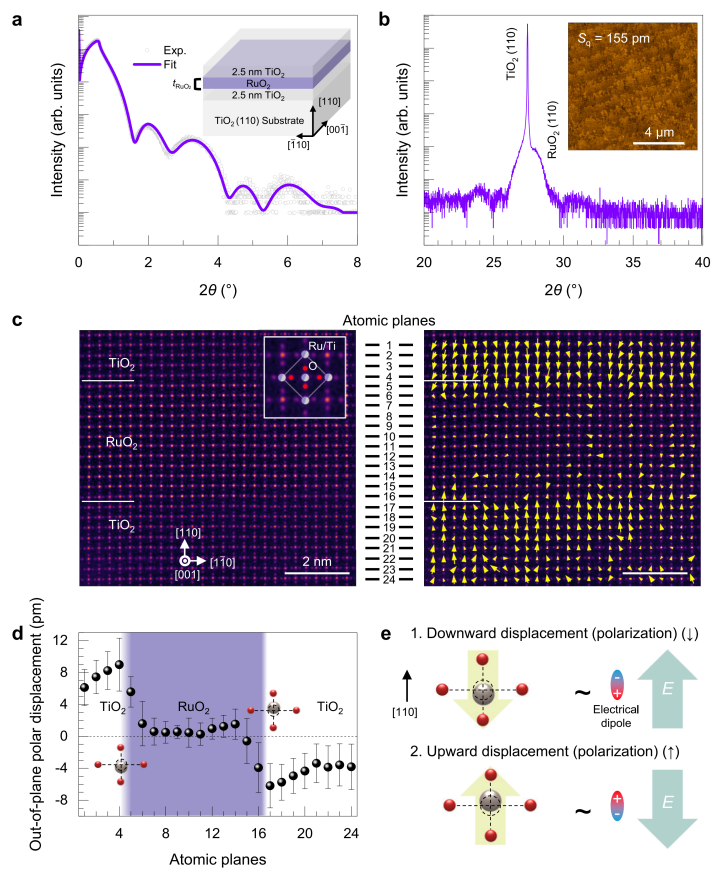

**Fig. 1. Interfacial polar displacement in TiO$_2$/RuO$_2$/TiO$_2$ (110) epitaxial heterostructure. a,b**, (**a**) XRR and (**b**) XRD $\theta$-$2\theta$ scan results for a 2.5 nm TiO$_2$/$t_{RuO_2}$ = 3.8 nm RuO$_2$/2.5 nm TiO$_2$ heterostructures, grown on the TiO$_2$ (110) substrate. The scattered symbols and lines in the XRR results indicate the experimental data and the corresponding fitting results, respectively. The inset in (**b**) shows an AFM image for the TiO$_2$/RuO$_2$/TiO$_2$ sample, demonstrating an atomically smooth surface with a step-terrace structure. **c**, Cross-sectional multislice electron ptychographic reconstruction (left panel) and polar displacement map (right panel) of 2.5 nm TiO$_2$/3.8 nm RuO$_2$/2.5 nm TiO$_2$ heterostructure. The inset in the left panel displays an overlay with the corresponding rutile structure. The largest arrow on the map indicates a polar displacement magnitude of 17.4 pm. **d,e**, (**d**) Out-of-plane polar displacement component averaged across each atomic plane and (**e**) a schematic illustration of polar displacements with two opposite out-of-plane directions and corresponding electrical dipoles.

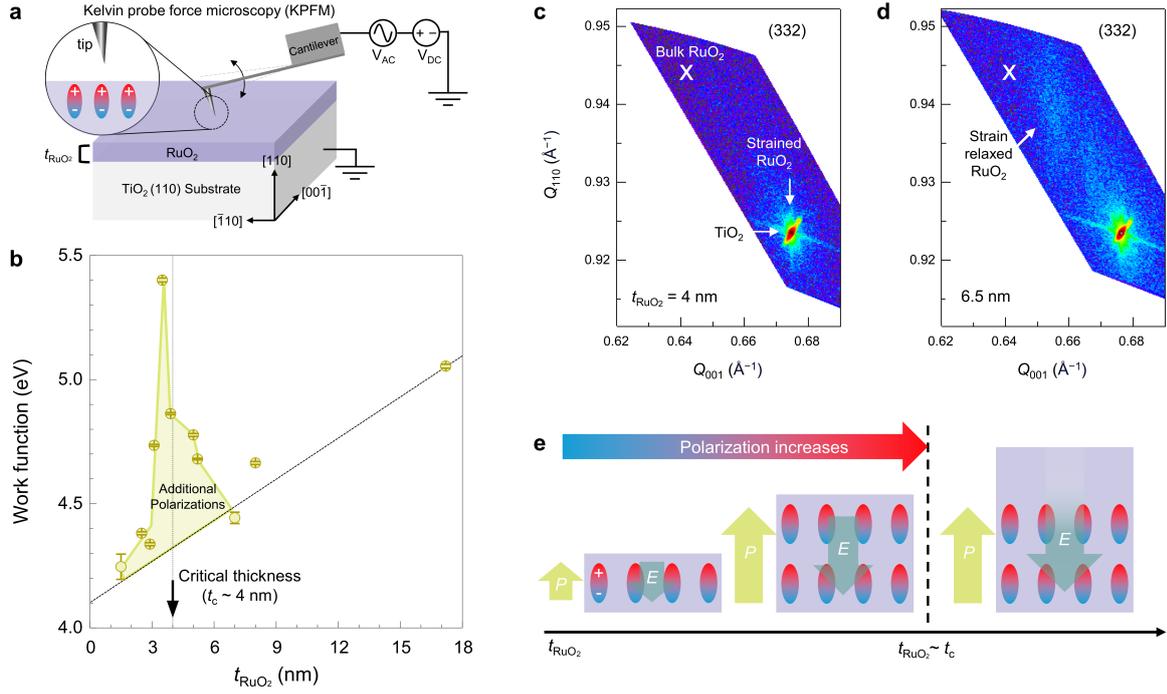

**Fig. 2. Interfacial polarization induced RuO$_2$ thickness evolutions of surface work function on RuO$_2$/TiO$_2$ (110) epitaxial thin films. a**, Schematic illustration of the KPFM setup with AC ($V_{AC}$) and DC voltages ($V_{DC}$) used to measure surface work function ($\Phi$) of epitaxial RuO$_2$/TiO$_2$ (110) thin films. **b**, $\Phi$ of RuO$_2$/TiO$_2$ (110) as a function of $t_{RuO_2}$, showing unexpectedly large enhancements. **c,d**, XRD RSM results around the (332) TiO$_2$ Bragg diffraction peak for epitaxial RuO$_2$/TiO$_2$ (110) thin films with (**c**) $t_{RuO_2}$ = 4 nm and (**d**) 6.5 nm. (**e**) Schematic illustration of the possible mechanism underlying the $\Phi$ variation in epitaxial RuO$_2$/TiO$_2$ (110) thin films. The polarization vector $P$ denotes the direction of atomic displacements, while the electric field vector $E$ represents the resulting internal electric field. A peak of $\Phi$ is observed near $t_{RuO_2} \sim t_c$, corresponding to the onset of strain relaxation, as identified in the XRD RSM results.

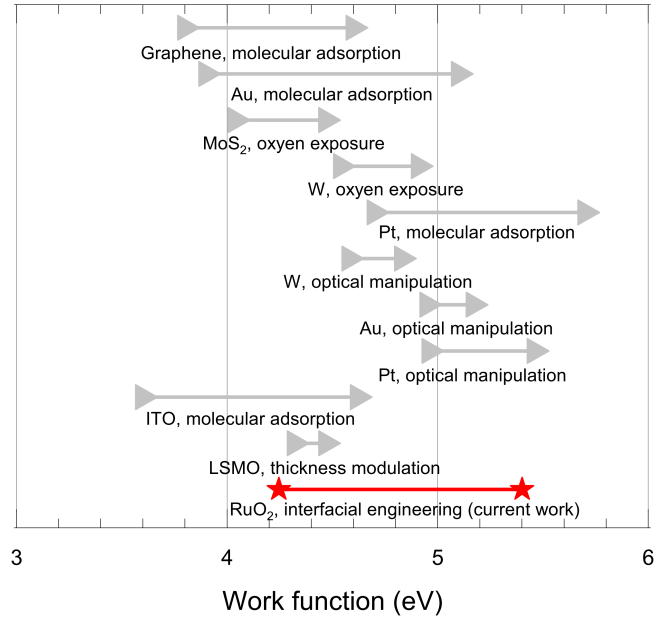

**Fig. 3. Comparison of work function modulations across various materials and methods.** The horizontal lines denote the modulation ranges between minimum and maximum work function values reported for each system: graphene (molecular adsorption[38]); Au (molecular adsorption[38] and optical manipulation[39]); $MoS_2$ (oxygen exposure[42]); W (oxygen exposure[40] and optical manipulation[39]); Pt (molecular adsorption[41] and optical manipulation[39]); ITO (Indium tin oxide, molecular adsorption[38]); LSMO ($La_{0.6}Sr_{0.4}MnO_3$, thickness modulation[37]); and $RuO_2$ (interfacial engineering, current work).

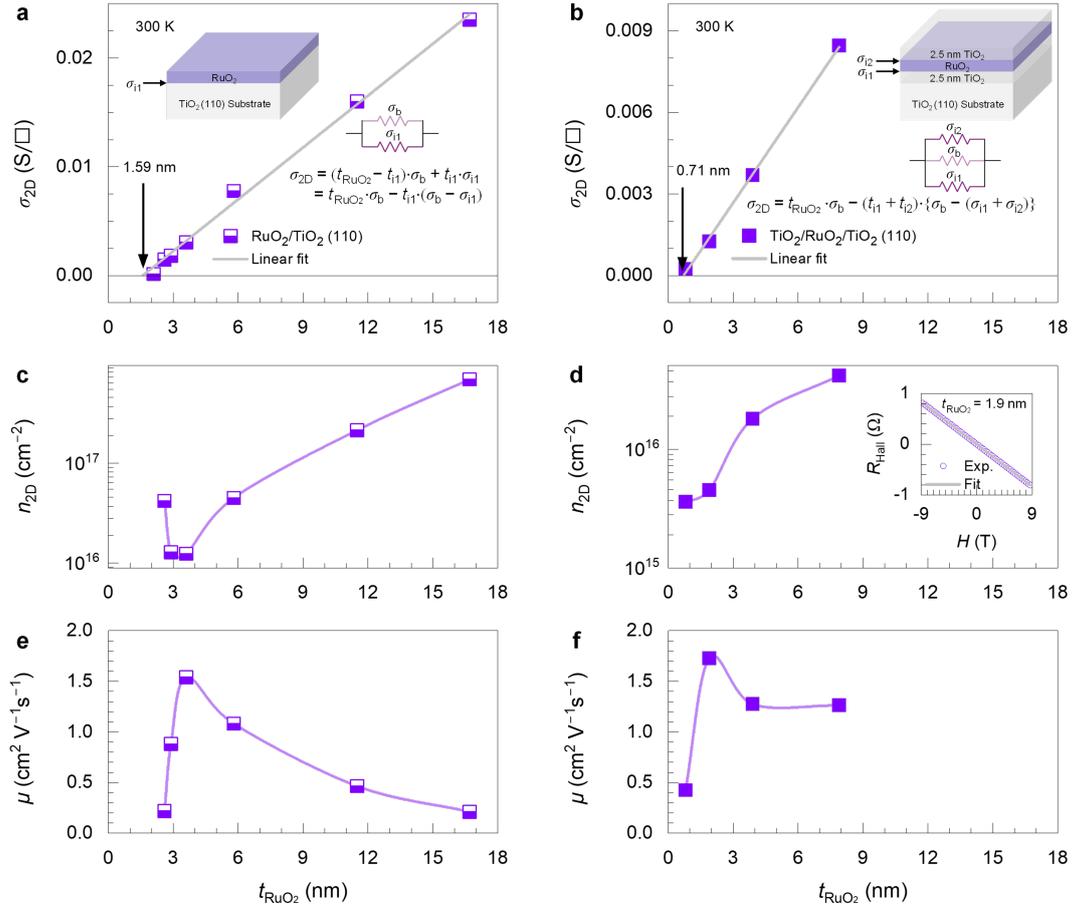

**Fig. 4. Modulation of electrical transport properties of polar metals at RuO$_2$/TiO$_2$ interfaces through epitaxial design. a,b,** 2D conductivity ($\sigma_{2D}$) as a function of $t_{RuO_2}$ (**a**) RuO$_2$/TiO$_2$ (110) films and (**b**) TiO$_2$/RuO$_2$/TiO$_2$ (110) heterostructure. The insets in (**a**) and (**b**) illustrate the corresponding heterostructure configurations and the two (or three) conductance models used to describe the $t_{RuO_2}$-dependent $\sigma_{2D}$. **c,d,** Sheet carrier density ($n_{2D}$) as a function of $t_{RuO_2}$ for (**c**) RuO$_2$/TiO$_2$ (110) films and (**d**) TiO$_2$/RuO$_2$/TiO$_2$ (110) heterostructures. The inset of (**d**) shows representative experimental data (scattered symbols) and fitting results (solid line) from the $H$-field-dependent Hall resistance ($R_{Hall}$) analysis for TiO$_2$/RuO$_2$/TiO$_2$ (110) heterostructures with $t_{RuO_2}$ = 1.9 nm. **e,f,** Carrier mobility ($\mu$) as a function of $t_{RuO_2}$ for (**e**) RuO$_2$/TiO$_2$ (110) films and (**f**) TiO$_2$/RuO$_2$/TiO$_2$ (110) heterostructures. The solid lines in **c,d,e,f** are a guide to the eye.

**Methods**

**Hybrid MBE and structural characterizations**

Epitaxial $RuO_2$ heterostructure films were grown on $TiO_2$ (110) single-crystal substrates (Crystec) using an oxide hybrid molecular beam epitaxy (hMBE) system (Scienta Omicron). The substrate preparation involved sequential cleaning with acetone, methanol, and isopropanol, followed by a 2-hour bake at 200 °C in a load-lock chamber. Prior to film growth, the substrates underwent a 20-minute annealing in oxygen plasma at 300 °C. For $RuO_2$ layer growth, a metal-organic precursor, $Ru(acac)_3$, was thermally evaporated using a low-temperature effusion cell (MBE Komponenten) operated between 170 and 180 °C. For $TiO_2$ layer growth, titanium was supplied using a liquid precursor, titanium tetraisopropoxide (TTIP, 99.999%, Sigma-Aldrich), which was injected into the MBE system via a line-of-sight gas injector (E-Science Inc.). The injection process was controlled using a customized gas-inlet system with a linear-leak valve and a Baratron capacitance manometer (MKS Instruments Inc.). The growth was carried out under a radio-frequency oxygen plasma at an oxygen pressure of $5 \times 10^{-6}$ Torr. To minimize the formation of oxygen vacancies, the sample was cooled to 120 °C in the presence of oxygen plasma after growth. The film surface evolution was monitored in situ using reflection high-energy electron diffraction (Staib Instruments) before, during, and after growth. Structural characterization—including crystallinity, film thickness, roughness, and strain state—was conducted using X-ray diffraction (Rigaku SmartLab XE) with reciprocal space mapping (RSM), X-ray reflectivity, and $\theta$-$2\theta$ measurements. In addition, surface morphology was analyzed using atomic force microscopy (Bruker Nanoscope V Multimode 8) in peak-force tapping mode.

**Kelvin probe force microscopy**

Kelvin Probe Force Microscopy (KPFM) was performed on an MFP-3D Infinity AFM (Asylum

Research, an Oxford Instrument branch, CA) using Pt-coated conductive AFM probes (Multi75E-G, BudgetSensors) in standard amplitude modulation mode. All KPFM measurements were conducted under ambient conditions, and the metallic samples were electrically grounded. The KFPM feedback tracks the DC voltage needed to minimize the 2$^{nd}$ harmonic vibration amplitude of the AFM cantilever electrically induced by the applied AC bias (Fig. 2a), which quantifies the contact potential difference ($V_{CPD}$) between the tip and sample surface. The sample work function $\Phi = \Phi_{tip} - qV_{CPD}$. The tip work function $\Phi_{tip}$ is calibrated by measuring the $V_{CPD}$ on a reference sample, 5.2 nm $RuO_2/TiO_2$ (110), whose work function is determined by ultraviolet photoelectron spectroscopy (UPS), as shown in Extended Data Fig. 6. Thus, the absolute values of $\Phi$ for various samples can be quantitatively determined and compared (standard error propagation were applied).

**Ultraviolet photoelectron spectroscopy**

Ultraviolet photoelectron spectroscopy (UPS) measurements were conducted using an Omicron XPS/UPS system. A DC bias was applied during the measurements to ensure the spectra would be within the detector range, which has been corrected in the presented data. The work function value (Fermi level is set at zero on the UPS spectra after correction) of the sample was calculated by $\Phi = h\nu - E_{sec}$, where $h\nu$ is the photon energy (21.2 eV for the He I ultraviolet source), and $E_{sec}$ is high binding energy cutoff (also known as secondary electron cutoff). The value of $E_{sec}$ was determined by identifying the intersection point between a tangent line fitted to the leading edge of the cutoff and the baseline (Extended Data Fig. 6).

**Electrical transport**

Electrical measurements were performed using the Van der Pauw geometry with aluminum wire bonding in a physical property measurement system (PPMS, Dynacool, Quantum Design, USA) with a temperature range between 1.8 K and 300 K, and a magnetic field range of ±9 T.

**Scanning transmission electron microscopy imaging and multislice electron ptychography**

Cross-sectional samples for electron microscopy were prepared along the [001] direction using a FEI Helios 600i DualBeam SEM/FIB, followed by $Ar^+$ ion milling. A Thermo Fisher Scientific Themis Z aberration-corrected scanning/transmission electron microscope operating at 200 kV was used for scanning transmission electron microscopy (STEM) imaging and collection of ptychographic 4D-STEM datasets. The beam current was set to 10-15 pA to minimize electron beam damage to the sample surface. For high-angle annular dark field (HAADF) STEM imaging, the incident probe had an 18.9 mrad convergence semi-angle, while the detector collection angle was 72-179 mrad. Image distortions from sample drift were corrected using the revolving STEM (revSTEM) method[45] on an 8-frame 1024×1024 image series, each frame with a 2 μs dwell time.

For acquiring ptychographic 4D-STEM datasets, the incident electron probe was overfocused by 10 nm and had a 26 mrad convergence semi-angle. The 4D-STEM dataset was collected using a 128 × 128 pixel Electron Microscope Pixel Array Detector (EMPAD)[46] with a diffraction pixel size of 0.78 mrad/px, a scan step size of 0.0443 nm/px, and a dwell time of 1 ms. Ptychographic reconstructions were performed using a modified version of the fold_slice[28] fork of the PtychoShelves[47] software package. The multislice reconstruction engine was a GPU-accelerated maximum-likelihood solver that utilized 16 incoherent probe modes [48, 49, 50]. In total, the reconstruction utilized two 300-iteration stages. In both stages, the depth regularization parameter was 0.1, and the probe was allowed to update from the first iteration. In the second stage, the diffraction patterns were zero-padded by a factor of two to double the real space sampling. The reconstructed sample thickness was 20 nm (1 nm slices), determined by adding several extra slices to the actual sample thickness of 12 nm estimated by the position-averaged convergent beam electron diffraction (PACBED)[51]. This allows the top and bottom slices to reconstruct the amorphous surface contamination and the electron beam-induced structure. The interior of the sample can be isolated from the surface by averaging only the three middle slices. Additionally, the impacts of sample tilt and surface relaxation are eliminated.

**Data availability**

The data that support the findings of this study are available within the Article and its Supplementary Information. Other relevant data are available from the corresponding authors upon reasonable request.

**Code availability**

The codes that support this study are available from the corresponding author upon reasonable request.

**Acknowledgements**

Film synthesis and structural characterizations (S.G.J. and B.J.) were supported by the U.S. Department of Energy through grant Nos. DE-SC0020211, and (partly) DE-SC0024710. Transport, and ellipsometry (at UMN) were supported by the Air Force Office of Scientific Research (AFOSR) through Grant No. FA9550-21-1-0025 and FA9550-24-1-0169. Film growth was performed using instrumentation funded by AFOSR DURIP awards FA9550-18-1-0294 and FA9550-23-1-0085. S.N. was supported partially by the UMN MRSEC program under Award No. DMR-2011401. Parts of this work were carried out at the Characterization Facility, University of Minnesota, which receives partial support from the NSF through the MRSEC program under Award No. DMR-2011401. M.J. and Q.T. acknowledge the funding support from the National Science Foundation under the Award No. CMMI-2311573 and the Texas A&M University System National Laboratory Office Seed Grant. Use of the TAMU Materials Characterization Facility (RRID: SCR_022202) is acknowledged. Parts of this work at GIST were supported by the National Research Foundation of Korea (NRF) grant funded by the Korea government (MSIT) (No. RS-2024-00486846). B.L. and J.M.L. acknowledge support from the AFOSR through grant No. FA9550-20-1-0066. 4D-STEM dataset collection was carried out using the facilities at MIT.nano, and ptychographic reconstructions were enabled by MIT SuperCloud and Lincoln Laboratory Supercomputing Center.

**Contributions**

S.G.J. and B.J. conceived the idea and established proof of concept. S.G.J., S.N., and A.S. grew films and characterized them using XRD, AFM, and electrical transport measurements. S.G.J. and B.J. analysed the data. B.L. and J.M.L. performed 4D-STEM dataset collection and ptychographic reconstructions. M.J. and Q.T. measured KPFM and UPS and analyzed data. B.J. directed and organized the different aspects of the project. S.G.J. and B.J. initiated writing the manuscript. All authors contributed to the discussion and preparation of the manuscript.

## Corresponding authors

Correspondence to Seung Gyo Jeong or Bharat Jalan.

## Competing interests

The authors declare no competing interests.